\documentclass[aps,prd,nofootinbib,twocolumn,floatfix,superscriptaddress,secnumarabic]{revtex4}
\pdfoutput=1
\usepackage{amsfonts,amssymb,epsfig,verbatim,mathbbol,amstext,amsmath,psfrag}
\usepackage{graphicx}
\usepackage[    bookmarks,
                 bookmarksopen = true,
                 bookmarksnumbered = true,
                 linktocpage,
                 colorlinks = true,
                 linkcolor = blue,
                 urlcolor  = blue,
                 citecolor = blue,
                 anchorcolor = green,
                 hyperindex = true,
                 hyperfigures]
		{hyperref}
\usepackage[latin2]{inputenc}
\usepackage{t1enc}
\usepackage{amsmath}
\usepackage{subfigure}
\usepackage{dsfont}
\usepackage{slashed}
\usepackage{multirow}
\usepackage{lscape}
\usepackage{wrapfig}
\usepackage{array}
\usepackage{mciteplus}

\usepackage{setspace}

\usepackage{colordvi}
\usepackage{lscape}
\usepackage{rotfloat}
\usepackage{rotating}
\usepackage{color}

\usepackage{cmap}
\newcommand{\be}{\begin{equation}}
\newcommand{\ee}{\end{equation}}

\newcommand{\Z}{\mathcal{Z}}
\renewcommand{\O}{\mathcal{O}}

\long\def\symbolfootnote[#1]#2{\begingroup%
\def\thefootnote{\fnsymbol{footnote}}\footnote[#1]{#2}\endgroup}

\def\dena{A(\mathcal{E})}
\def\dbna{A(\mathcal{B})}
\def\F{\mathcal{C}}
\newcommand{\FF}{F^{\rm ps}}
\newcommand{\PP}{p^{\rm ps}}

\begin{document}
\title{Paramagnetic squeezing of QCD matter}

\author{G.~S.~Bali}
\affiliation{Institute for Theoretical Physics, Universit\"at Regensburg, D-93040 Regensburg, Germany}
\affiliation{Department of Theoretical Physics, Tata Institute of Fundamental Research, Homi Bhabha Road, Mumbai 400005, India.}
\author{F.~Bruckmann}
\affiliation{Institute for Theoretical Physics, Universit\"at Regensburg, D-93040 Regensburg, Germany}
\author{G.~Endr\H{o}di}
\email[Corresponding author, email: ]{gergely.endrodi@physik.uni-r.de}
\affiliation{Institute for Theoretical Physics, Universit\"at Regensburg, D-93040 Regensburg, Germany}
\author{A.~Sch\"afer}
\affiliation{Institute for Theoretical Physics, Universit\"at Regensburg, D-93040 Regensburg, Germany}

\begin{abstract}
We determine the magnetization of Quantum Chromodynamics (QCD)
for several temperatures around and above the transition 
between the hadronic and the quark-gluon phases of strongly
interacting matter. We obtain
a paramagnetic response that increases in strength with
the temperature. We argue that due to this paramagnetism,
chunks of quark-gluon plasma produced in non-central heavy ion
collisions should become elongated along the direction
of the magnetic field.
This anisotropy will then contribute to the elliptic flow $v_2$
observed in such collisions, in addition to the pressure gradient
that is usually taken into account.
We present a simple estimate for the magnitude of this new effect and 
a rough comparison to the effect due to the initial collision geometry.
We conclude that the paramagnetic effect might have a significant impact on the value of $v_2$. 
\end{abstract}

\pacs{12.38.Gc, 12.38.Mh, 25.75.Ld, 25.75.Ag}
\keywords{lattice QCD, quark-gluon plasma, external fields, heavy-ion collision, elliptic flow}

\maketitle

\section{Introduction}

In heavy-ion collisions (HICs) strongly interacting
matter is exposed to extreme conditions to
probe the QCD phase diagram and to reveal properties
of the quark-gluon plasma (QGP). It is however not straightforward to
relate characteristics of the so produced QCD medium to
experimental signatures. 
One of the most prominent experimental
observables is the elliptic flow $v_2$~\cite{Csernai:1999nf},
which marks the onset of hydrodynamic behavior at very early times 
(hydroization). 
Connecting $v_2$ to the centrality of HICs in a model-independent 
way is crucial to extract the ratio 
of viscosity to entropy density $\eta/s$ of the QGP~\cite{Romatschke:2007mq}.

Another important aspect of the initial phase of HICs is the 
generation of extremely strong magnetic 
fields~\cite{Skokov:2009qp,Bzdak:2011yy,Voronyuk:2011jd,Deng:2012pc}.
We show that these magnetic fields may have
an impact on $v_2$ and, therefore, should be taken into
account in a quantitative analysis of the elliptic flow.
Irrespective of this observable effect, the response to
magnetic fields is a fundamental property of QCD matter
which deserves to be studied in its own right.
Other applications of our findings 
include models  of neutron stars (magnetars~\cite{Duncan:1992hi})
and primordial magnetic fields in the early 
universe (see, e.g., Ref.~\cite{Grasso:2000wj}).

All information about the response of QCD to magnetic fields
can be deduced from the free energy density
$f= -T/V\cdot\log\Z$,
given in terms of the partition function $\Z$. 
Applying a constant external magnetic field $B$
induces a nonzero magnetization
\be
M= - \frac{\partial f}{\partial (eB)},
\label{eq:Mdef}
\ee
which we normalized by the elementary charge ($e>0$). 
The sign of $M$ determines whether the QCD vacuum as a medium
exhibits a \textit{para}magnetic response 
($M>0$) or a \textit{dia}magnetic one ($M<0$)~\cite{landau1995electrodynamics}. 
In the former case the magnetization is aligned parallel
to the external field, while in the latter case it is antiparallel.
One clue about the sign of $M$ came from a low-energy
effective model of QCD
--- the hadron resonance gas (HRG) model ---
which predicted the magnetization to be positive 
and thus the QCD vacuum to be a paramagnet~\cite{Endrodi:2013cs}. 
Several methods were since developed to study the problem on the
lattice~\cite{Bali:2013esa,Bonati:2013lca,*Bonati:2013vba,Levkova:2013qda,Bali:2013txa}. 
All of the results agree qualitatively, confirming the 
finding of the HRG model that the QCD vacuum is {\it para}magnetic.

In the present letter we
extend the lattice measurements of Ref.~\cite{Bali:2013esa}
to cover several temperatures in and above the transition region. 
We do not yet provide final,
continuum extrapolated values for the
magnetization, but instead aim at a first estimate of the effect
of QCD paramagnetism on the phenomenology of heavy ion
collisions.
\begin{figure}[t]
\centering
\vspace*{-0.3cm}
\includegraphics[width=6.0cm]{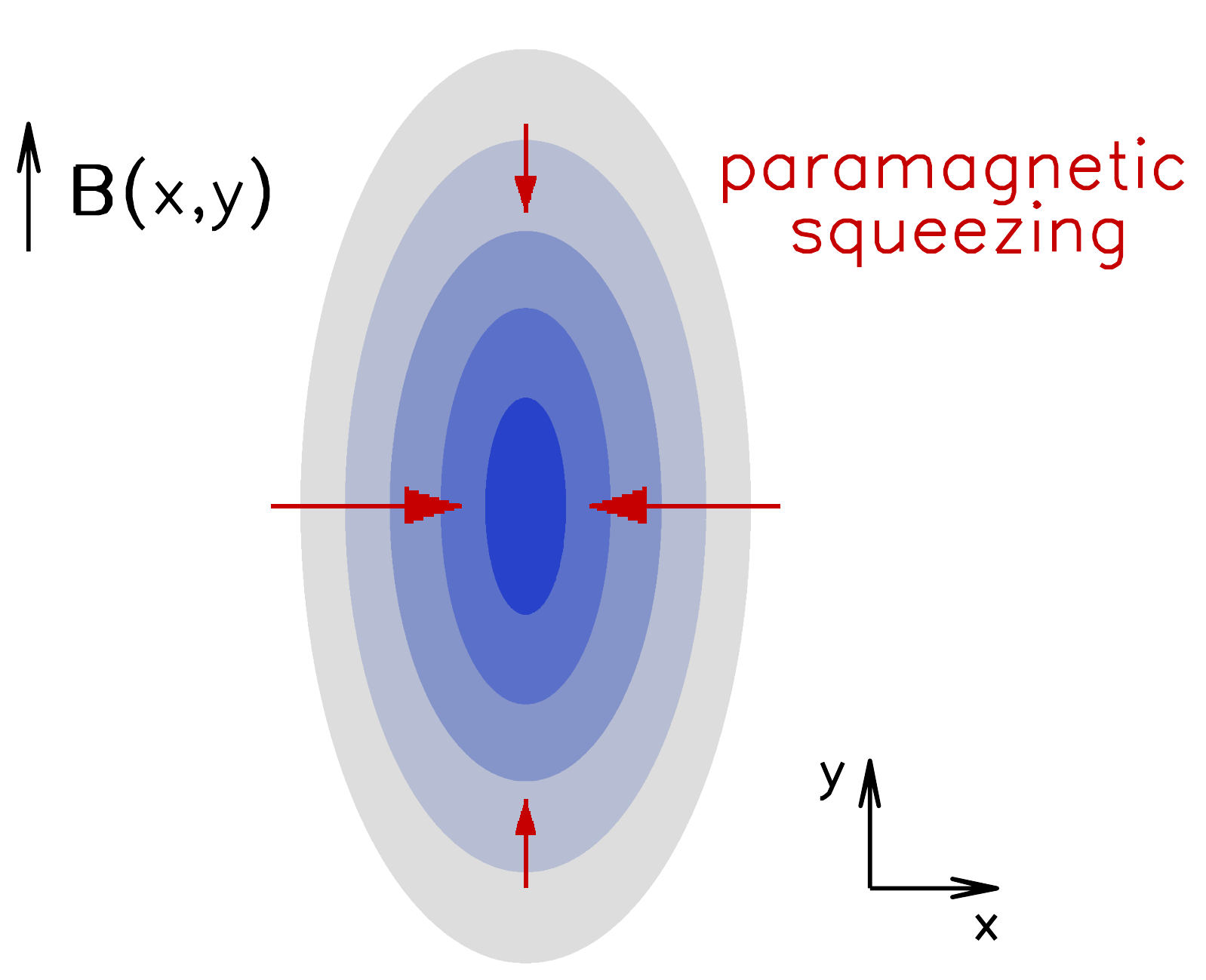}
\vspace*{-0.2cm}
\caption{\label{fig:v2}
Typical magnetic field profile in the transverse plane of 
a non-central heavy-ion collision (darker colors represent stronger fields). 
The paramagnetic squeezing exerts the force indicated by the red arrows. 
As a result, the QGP is elongated in the $y$-direction.}
\vspace*{-.2cm}
\end{figure}
To this end, let us consider a chunk of the QGP exposed to a non-uniform magnetic 
field. Owing to the positivity of $M$, the free energy is minimized when the 
medium is located in regions where $B$ is maximal. 
The minimization of $f$ thus results in
a net force, which strives to change the shape of the medium.
For a non-central HIC
(with $\hat z$ being the direction of the collision axis, 
$\hat x$-$\hat z$ the reaction plane and 
$\hat y$ the direction of the magnetic field induced by the beams),
this force will elongate the distribution of QCD matter along the
$y$-direction, see Fig.~\ref{fig:v2}. 
This elongation can affect the azimuthal structure of the expansion 
of the system, in addition to the pressure gradients due to
the initial geometry.
We call this effect {\it paramagnetic squeezing}.

The interplay of the geometric pressure gradient and
the paramagnetic squeezing crucially depends
both on the time- and space-dependence of the magnetic field 
\cite{Tuchin:2013apa,McLerran:2013hla}
and on the moment of the onset of early
hydroization~\cite{Heller:2011ju,*vanderSchee:2013pia}. 
Both time scales are subjects of
ongoing debate. We also remark that our quantitative results
apply, strictly speaking, to QCD matter in thermal equilibrium only.
Therefore, at present the best we can do is to estimate whether the
impact of the sketched squeezing effect is sizeable.
According to our numerical results, which we detail below,
this can be the case for RHIC and in particular for LHC collisions.
This clearly calls for a quantitative analysis
of the dynamics of the decay of the initial magnetic field and
of the on-set of hydroization.

We mention that the paramagnetic squeezing may correlate
the recently observed large event-by-event fluctuations in $v_2$~\cite{Timmins:2013hq} 
with fluctuations of the magnetic field.
Furthermore, we note that a similar squeezing effect may operate 
in the inner core of magnetars if the magnetic field there reaches typical QCD scales ($B\sim 10^{14-15}\textmd{ T}$).

\section{Lattice setup}

We consider a hypercubic lattice of size $N_s^3\times N_t$ and spacing $a$.
The discretization of the QCD action we choose is the tree-level Symanzik improved 
gluonic action and stout smeared staggered quarks for the three lightest flavors 
(the detailed simulation setup is described in Refs.~\cite{Aoki:2005vt,Bali:2011qj}).
The quark masses are set to their physical values along the line of constant physics 
(for details see Ref.~\cite{Borsanyi:2010cj}), $m_u=m_d=m_s/28.15$. 
The electric charges of the quarks are
$q_u/2=-q_d=-q_s=e/3$. The flux $\Phi$ of the magnetic field on the lattice is quantized,
\be
\Phi \equiv (N_sa)^2 \cdot e B = 6\pi N_b,\quad N_b\in \mathds{Z},\quad0\le N_b<N_s^2,
\label{eq:quant}
\ee
which prohibits 
the direct evaluation of the magnetization, Eq.~(\ref{eq:Mdef}),
as a derivative with respect to $B$. 
The first approach to circumvent this problem 
was developed in Ref.~\cite{Bali:2013esa}, where we calculated $M$
as the difference of lattice pressures parallel and perpendicular
to $B$. 
Several alternatives were also introduced recently to determine the 
magnetization. In Refs.~\cite{Bonati:2013lca,*Bonati:2013vba}, the derivative of $f$ 
with respect to $B$ is constructed (giving an unphysical quantity due to flux quantization) 
and then integrated to obtain the physical change of the free energy due to $B$.
Ref.~\cite{Levkova:2013qda} evades flux quantization 
altogether by considering a magnetic field which is positive in one and 
negative in the other half of the lattice. 
Finally, in Ref.~\cite{Bali:2013txa} we developed an integral method which is based 
on the $B$-independence of $f$ at asymptotically large quark masses.

Here we follow the approach of Ref.~\cite{Bali:2013esa}. 
The main idea is the following: Flux quantization implies that a
hypothetical compression of the 
system in the direction perpendicular to $B$ can only proceed
keeping $\Phi$ fixed -- which automatically implies changing $B$. Therefore, one has
to compress the magnetic 
field lines together with the system. In Ref.~\cite{Bali:2013esa}
we have shown that this 
gives a consistency relation between the response of the system to a change in 
$B$ -- i.e., 
the magnetization -- and the response to a change in the size of the system
in different directions -- i.e. the pressures. This leads to the relation
\be
-M\cdot eB = -(\zeta_g+\hat\zeta_g) \cdot [\dbna-\dena] - \zeta_f \cdot \sum_f A(\F_f),
\ee
where $\dbna$, $\dena$ are the anisotropies in the chromomagnetic
and chromoelectric parts
of the gluonic action, $A(\F_f)$ is the anisotropy in the
fermionic action for the 
quark flavor $f$ ($f=u,d,s$), and $\zeta_g$, $\hat\zeta_g$ and $\zeta_f$ are 
renormalization coefficients that may be determined, e.g., by simulating on 
anisotropic lattices. To leading
order in perturbation theory, $\zeta_g=\hat\zeta_g=\zeta_f=1$.
We found the gluonic anisotropies to be by factors of 5--10 smaller than the 
fermionic anisotropy. This means that the magnetization is well
approximated by
$M\cdot eB \approx \sum_f A(\F_f)$, 
within a systematic error of 10--20~\%. The fermionic anisotropy in terms 
of the parallel and perpendicular components of the Dirac operator reads
\be
A(\F_f) = \frac{1}{2} \left[ \bar\Psi_f \slashed{D}_x \Psi_f + \bar\Psi_f \slashed{D}_y \Psi_f \right] 
-  \bar\Psi_f \slashed{D}_z \Psi_f,
\ee
where $\slashed{D}_\mu$ is the component of the Dirac operator proportional to $\gamma_\mu$, which 
is readily accessible on the lattice.

\section{Renormalization}

\begin{figure}[b]
\centering
\vspace*{-.3cm}
\includegraphics[width=6.5cm]{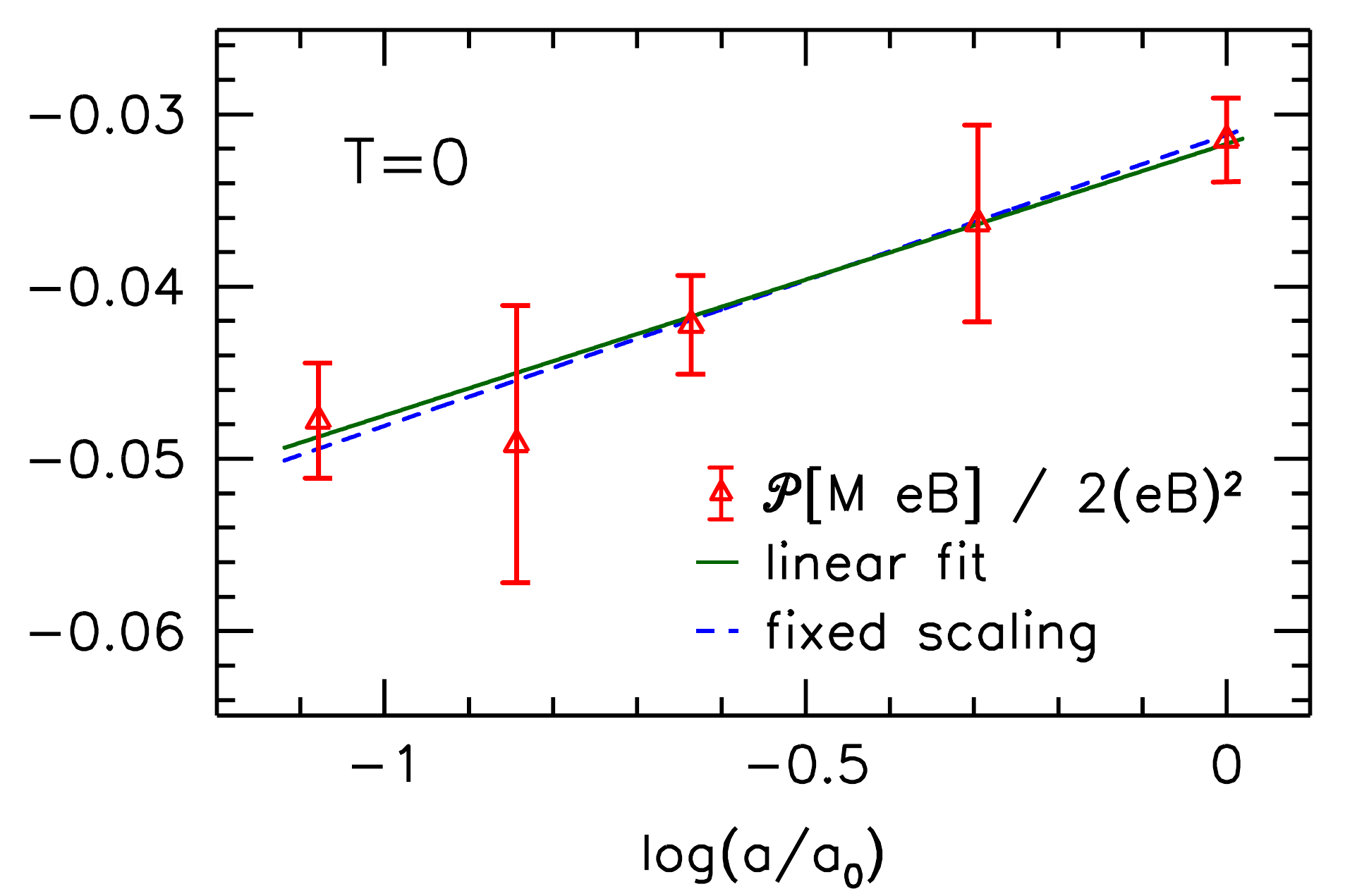}
\vspace*{-.3cm}
\caption{\label{fig:Mrenorm}
The divergent $\mathcal{O}((eB)^2)$ contribution to the magnetization 
normalized by $2(eB)^2$, as a function 
of the logarithm of the lattice spacing
(in units of $a_0=1.47\textmd{ GeV}^{-1}$, 
our coarsest lattice). 
A linear fit with two free parameters (green solid line) and one with 
the slope fixed to the leading perturbative prediction $\beta_1^{\rm QED}$ (see text) is also shown
(dashed blue line).}
\vspace*{-.1cm}
\end{figure}

\begin{figure*}[ht!]
\centering
\includegraphics[width=17cm]{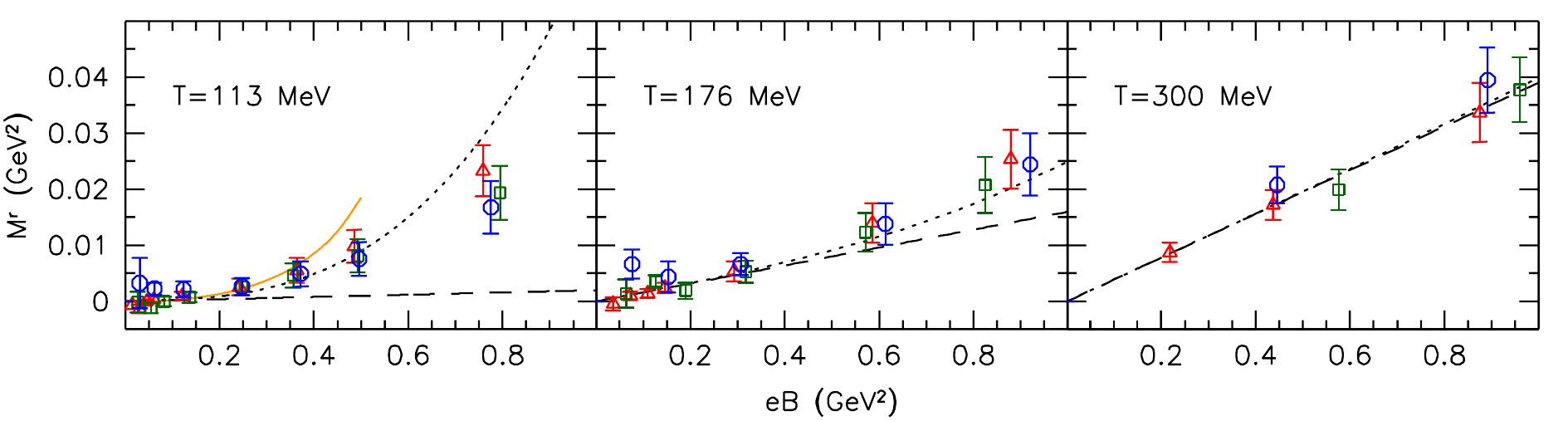}
\vspace*{-.3cm}
\caption{\label{fig:magnT}
Renormalized magnetization of QCD as a function of the magnetic field
for various values of temperature.
Different colors encode different lattice spacings: 
$N_t=6$ (red triangles), $N_t=8$ (green squares) and $N_t=10$ (green circles). The 
continuum limit corresponds to $N_t\to\infty$. Below $T_c$ we show the HRG model 
prediction~\protect\cite{Endrodi:2013cs} (solid orange line). 
The dashed lines represent the linear terms in $M^r$, which were determined by fitting 
the magnetization at small magnetic fields, whereas the dotted curves show the results 
of cubic fits.
}
\vspace*{-.1cm}
\end{figure*}

The free energy in the presence of an external magnetic field contains a $B$-dependent 
logarithmic divergence. This divergence stems from the coupling of quarks to $B$ through 
their electric charges and is cancelled by the renormalization of the electric charge~\cite{Schwinger:1951nm}. 
The fact that the prefactor of the divergence in 
question is given by the lowest-order QED $\beta$-function
coefficient $\beta_1$
(this contains perturbative and non-perturbative QCD corrections)
further illustrates the fundamental relation between the $B$-dependent
divergence of $f$ and electric charge 
renormalization.
At zero temperature, this divergence constitutes the only term which is of $\mathcal{O}((eB)^2)$:
\be
f(B)-f(0) = -\beta_1 \cdot (eB)^2 \cdot \log(a) + \mathcal{O}((eB)^4).
\ee

The magnetization inherits this divergence and -- again, at zero temperature -- 
has the structure 
\be
M \cdot eB = 2\beta_1 \cdot (eB)^2 \cdot \log(a) + \mathcal{O}((eB)^4).
\label{eq:Mrenorm}
\ee
The renormalization of $M$ therefore amounts to subtracting the total 
$\O((eB)^2)$ term at zero temperature (for a more detailed explanation of
this point 
see, e.g., Refs.~\cite{Elmfors:1993bm,*Dunne:2004nc,Endrodi:2013cs}). This term can be 
determined by considering the limit of small magnetic fields.
This results in the 
renormalization prescription
\be
f^r = (1-\mathcal{P}) [f], \quad\quad
M^{r} \cdot eB = (1-\mathcal{P}) [M\cdot eB],
\label{eq:Mrdef}
\ee
where $\mathcal{P}$ is
the operator that projects out the $\mathcal{O}((eB)^2)$ term from  an observable $X$:
\be
\mathcal{P}[X] = (eB)^2 \cdot \lim_{eB\to0} \frac{X}{(eB)^2}.
\ee

In Fig.~\ref{fig:Mrenorm} we show the coefficient of the divergent term $\mathcal{P}[M\cdot eB]/2(eB)^2$ 
for several lattice spacings at $T=0$. 
In accordance with Eq.~(\ref{eq:Mrenorm}), the divergent term is found to be proportional 
to $\log(a)$, with a coefficient of $0.016(4)$. The leading order perturbative scaling is 
given by the lowest order coefficient of the QED $\beta$-function for three quark flavors
with $N_c=3$ colors, 
$\beta_1^{\mathrm{QED}} = N_c/(12\pi^2) \cdot \sum_{f=u,d,s} (q_f/e)^2 \approx 0.0169$:
the fitted slope is consistent with the perturbative prediction, within 
statistical errors. 
Increasing the statistics would be necessary to resolve 
QCD corrections to $\beta_1$.

\section{Results and discussion}

We can now subtract the divergent part of $M$ measured at $T=0$ 
to obtain the temperature-dependence of the renormalized magnetization $M^r$, as defined 
in Eq.~(\ref{eq:Mrdef}). 
The result is shown in Fig.~\ref{fig:magnT}, where $M^r$ is 
plotted for $eB<1.0 \textmd{ GeV}^2$ at three values of the temperature.
The results for all three lattice spacings fall essentially on 
top of each other, indicating small lattice artefacts.
(Note that the renormalization at $T=300 \textmd{ MeV}$, where the lattice 
spacing is the smallest, requires an
extrapolation of the $T=0$ contribution of Fig.~\ref{fig:Mrenorm}. 
The systematic error due to this extrapolation 
is taken into account here.)

We find $M^r>0$ for all temperatures, which demonstrates 
the paramagnetic nature of the thermal QCD vacuum. This is in agreement with our earlier 
results at $T=0$~\cite{Bali:2013esa}. Note that the expansion of $M^r$ in the magnetic field 
at $T=0$ starts as $(eB)^3$. As the temperature increases,
thermal contributions 
induce an additional linear term in $M^r$, as is visible in the plots (shown by the dashed lines). 
This behavior is also present in the HRG model prediction,
which we include in the 
figure for the lowest temperature,
where the hadronic description is expected to 
be still valid. We find the HRG model to reproduce the
lattice data for small 
fields $eB\lesssim 0.3 \textmd{ GeV}^2$. 

The results are well described by a cubic function 
$M^r=\chi_1\cdot eB + \chi_3\cdot(eB)^3/T^4$, shown by the dotted lines in 
Fig.~\ref{fig:magnT}. 
The linear coefficient is the magnetic susceptibility, 
marking the leading response of QCD to the external field. 
It is zero at low temperatures and increases drastically above the transition. 
On the other hand, the cubic term is found to decay strongly with temperature, 
consistent with $T^{-4}$ as expected on dimensional grounds. 
The fit parameters are listed in Table~\ref{tab:1}. 
\setlength{\tabcolsep}{3pt}
\setlength{\extrarowheight}{1.5pt}
\begin{table}[h]
 \begin{tabular}{c|c|c|c|c|c}
  $T$ (MeV) & 113 & 130 & 142 & 176 & 300 \\ \hline
  $\chi_1\cdot 10^2$ & 0.2(5) & 0.4(7) & 1.1(6) & 1.6(5) & 3.9(6) \\
  $\chi_3 \cdot 10^6$ & 9(5) & 8(3) & 6(3) & 9(5) & 7(5) \\
 \end{tabular}
 \caption{\label{tab:1}Temperature-dependence of the coefficients $\chi_1$ and $\chi_3$.
To obtain the (linear) magnetic susceptibility in SI units, 
one needs to multiply $\chi_1$ by $e^2 \approx 4\pi/137$.}
\end{table}
We remark that for high temperatures we expect $\chi_1$ 
to show a logarithmic rise with $T$; the Stefan-Boltzmann limit 
of $\chi_1$ for free massive quarks is $2\beta_1^{\rm QED}\cdot \log(T/m)$~\cite{Elmfors:1993wj}.
(Note that such an entanglement between ultraviolet and infrared divergences for nonzero 
magnetic fields is well-known, see, e.g., Ref.~\cite{Dunne:2002ta}.)
The high-temperature limit of the next coefficient is 
$\chi_3 = 31N_c\,\zeta(5)/(960\pi^6) \cdot \sum_{f=u,d,s} (q_f/e)^4
\approx 2.32\cdot 10^{-5}$, with a negative $m^2/T^2$ correction~\cite{Cangemi:1996tp}. 
The lattice data indeed approaches this limit from below.

\begin{figure}[t]
\centering
\hspace*{-.1cm}
\mbox{
\includegraphics[width=4.16cm]{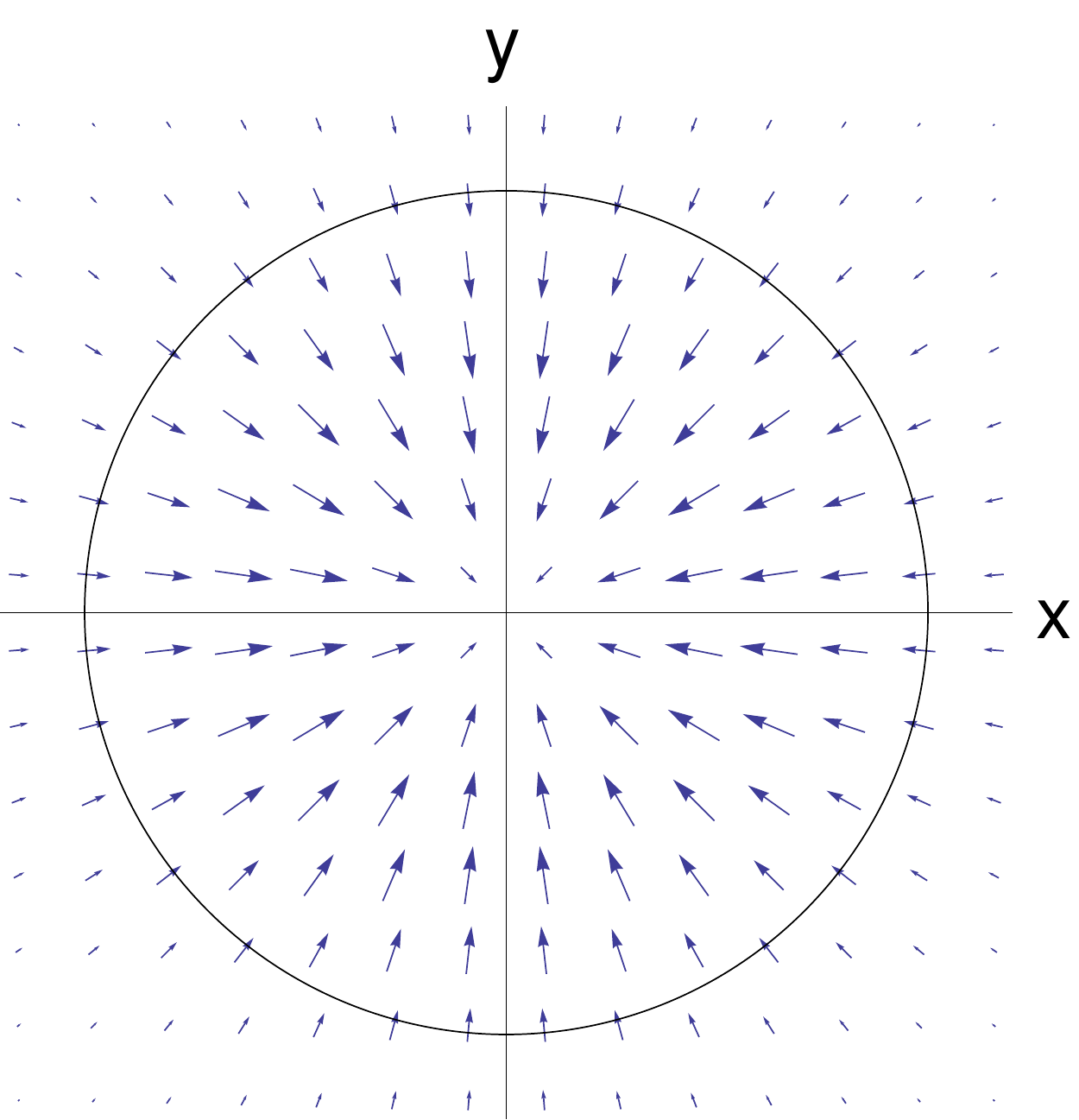}
\includegraphics[width=4.16cm]{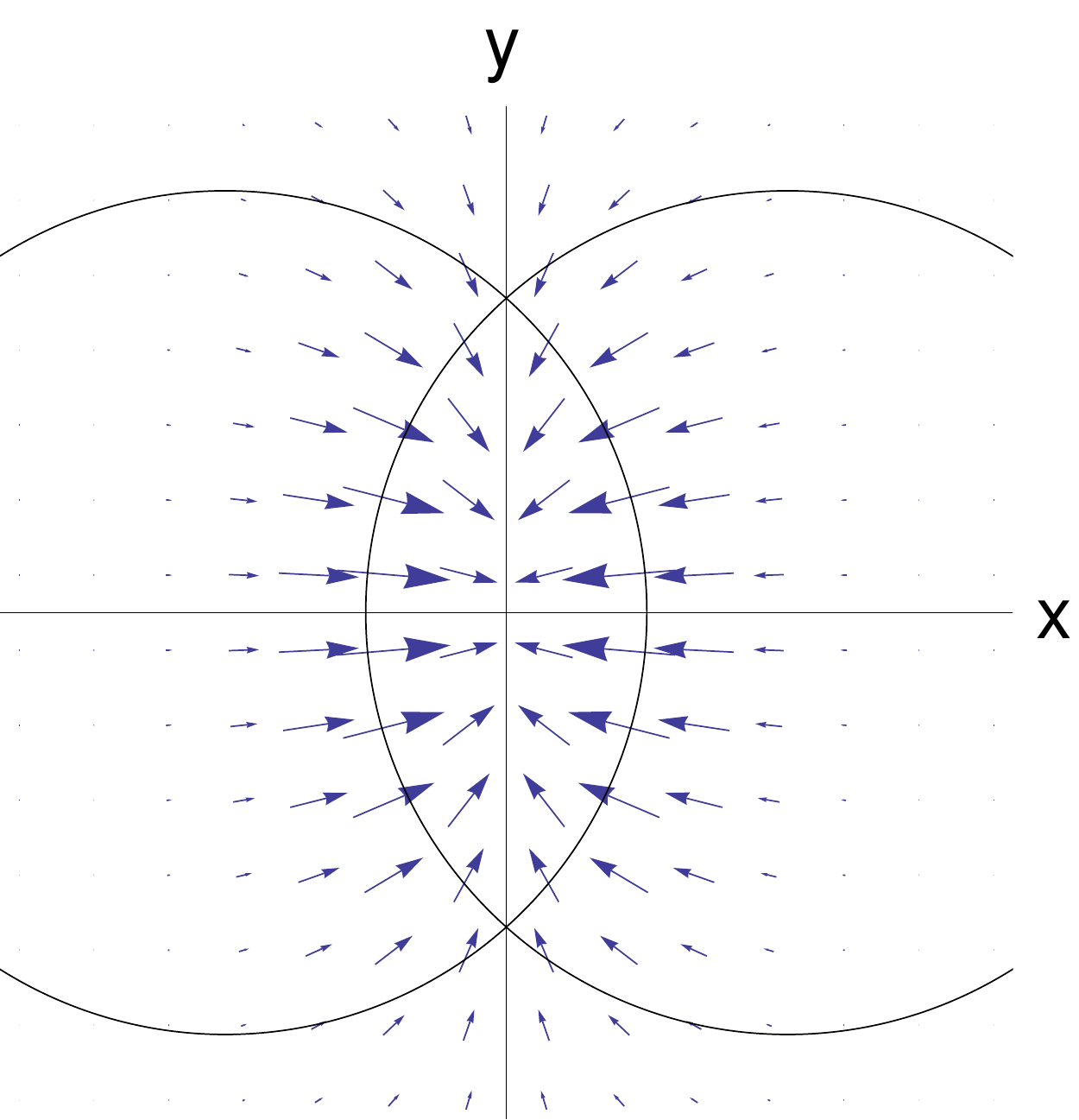}
}
\caption{\label{fig:illustr}Paramagnetic force profiles for typical central (left panel) 
and peripheral (right panel) heavy-ion collisions. The circles represent the colliding nuclei. 
}
\end{figure}

We are now in the position to discuss the implications of our results 
on HIC phenomenology. 
For a non-uniform magnetic field, the paramagnetic squeezing is manifested by a 
force density $\FF$, which arises as the system strives to minimize its free energy,
\be
\FF = -\nabla f^r = - \frac{\partial f^r}{\partial (eB)} \cdot \mathbf{\nabla} (eB) = M^r \cdot \mathbf{\nabla} |eB|.
\label{eq:Fpm}
\ee
Note that the free energy density -- being a Lorentz-scalar -- 
is only sensitive to the magnitude of the field, which allowed us to replace $eB$ 
by $|eB|$ above. Motivated by model descriptions of the magnetic field 
profile~\cite{Voronyuk:2011jd,Deng:2012pc}, 
we consider a simple two-dimensional Gaussian distribution 
of the magnetic field (with widths $\sigma_x=\sigma_y$ for a central collision 
and $\sigma_x = \sigma_y/2$ for a peripheral collision). 
The so obtained force profiles are depicted in Fig.~\ref{fig:illustr}.
While isotropic for central collisions, this inward-pointing 
force becomes anisotropic for the peripheral case, squeezing and hence
elongating the medium distribution.

At high temperatures the magnetization is linear in $eB$ (see the right panel of 
Fig.~\ref{fig:magnT}), and the gradient also contains a factor of $eB$, 
making $\FF$ proportional to the square of the magnetic field. 
Accordingly, the effect is very sensitive to the spatial profile of the 
magnetic field generated in the collision. Furthermore, this profile 
depends strongly on the time elapsed since the moment of the 
collision~\cite{Tuchin:2013apa}, giving rise to a complex behavior of 
the squeezing effect as a function of space and time.

We now make a first attempt to estimate the strength of the paramagnetic
squeezing effect based on very simplistic assumptions, and quantify
it in terms of the difference between the magnitudes of the force densities acting at $(\sigma_x,0)$ 
and at $(0,\sigma_y)$. This is equivalent to a 
difference $\Delta \PP$ of pressure gradients. 
From our results for the magnetization at $T=300\textmd{ MeV}$ and the magnetic 
field profiles of Refs.~\cite{Voronyuk:2011jd,Deng:2012pc}, we obtain 
$|\Delta \PP| \approx 0.007 \textmd{ GeV}/\textmd{fm}^4$ for magnetic fields 
of the order of $5 \, m_\pi^2$ (the typical 
value obtained in model calculations for RHIC 
energies~\cite{Voronyuk:2011jd,Deng:2012pc,Bloczynski:2012en}).
Using the magnetic field $50 \,m_\pi^2$ 
corresponding to LHC energies~\cite{Deng:2012pc} amounts to 
$|\Delta \PP| \approx 0.7 \textmd{ GeV}/\textmd{fm}^4$. 
Similar estimates for the early-time pressure gradients $p^{\rm g}$ resulting from the 
geometric effect give differences of 
$|\Delta p^{\rm g}| \approx 0.1 \textmd{ GeV}/\textmd{fm}^4$ for RHIC 
and $|\Delta p^{\rm g}| \approx 1\textmd{ GeV}/\textmd{fm}^4$ for
LHC energies~\cite{Pasi_correspondence}, see also 
Refs.~\cite{Kolb:2000sd,*Petersen:2006vm}. 
The estimates for the paramagnetic squeezing 
are subject to large systematic errors 
originating, for example, from the uncertainty of the QGP electric 
conductivity~\cite{Tuchin:2013apa,McLerran:2013hla}. 
Moreover, due to the complex space- and time-dependence of both mechanisms, 
a comparison based only on the magnitude of $\FF$ at two spatial points is clearly 
too simplistic. Instead, one should consider a model for the early stage ($t\lesssim1\textmd{ fm}/c$) 
of the collision, and take the paramagnetic squeezing effect into account from 
the beginning. This is outside the scope of the present letter.

To summarize: we have studied the magnetic response of the quark-gluon 
plasma by means of lattice simulations at physical quark masses. The response 
is {\it paramagnetic} and the magnetization at different 
temperatures is plotted in Fig.~\ref{fig:magnT} and parameterized in Table~\ref{tab:1}. 
Based on these equilibrium results we estimated the {\it paramagnetic squeezing} effect 
on chunks of quark-gluon plasma produced in heavy ion collisions.
We compared its magnitude 
to pressure gradients arising from geometrical effects.
The energy-dependence of the two mechanisms is quite different. 
For typical HICs at RHIC we found the paramagnetic squeezing to
give a $10\%$ correction to the geometric effect, while 
for LHC collisions they are similar in size. 
Our estimates should be improved by more involved model calculations
of the elliptic flow, taking into account the
paramagnetic squeezing in the evolution of the fireball. 
As both the onset of hydroization and the 
magnetic field dynamics are still subject of debates,
this calls for a dedicated large scale effort. 

\noindent {\bf Acknowledgements.} 
Our work was supported by the DFG (SFB/TRR 55), the EU (ITN STRONGnet
238353) and the Alexander von Humboldt Foundation. 
The authors thank Pasi Huovinen, Simon Mages, 
Berndt M\"{u}ller and Hannah Petersen for useful discussions.

\bibliographystyle{JHEP_mcite}
\bibliography{elliptic}

\end{document}